\documentstyle[aps,prl,epsf,floats]{revtex}
\begin{document}
\twocolumn[\hsize\textwidth\columnwidth\hsize\csname@twocolumnfalse%
\endcsname

\title{Magnetic properties and spin waves of bilayer magnets in a uniform field}

\author{Torsten Sommer$^{(a)}$, Matthias Vojta$^{(b)}$, and Klaus~W. Becker$^{(a)}$}
\address{(a) Institut f\"{u}r Theoretische Physik, Technische Universit\"{a}t Dresden,
D-01062 Dresden, Germany\\
(b) Theoretische Physik III, Elektronische Korrelationen und
Magnetismus, Institut f\"ur Physik,\\ Universit\"at Augsburg,
D-86135 Augsburg, Germany}
\date{August 5, 2001}
\maketitle

\begin{abstract}

The two-layer square lattice quantum antiferromagnet with spins $1 \over 2$ shows a
zero-field magnetic order-disorder transition at a critical ratio of the inter-plane to
intra-plane couplings.
Adding a uniform magnetic field tunes the system to canted antiferromagnetism and
eventually to a fully polarized state;
similar behavior occurs for ferromagnetic intra-plane coupling.
Based on a bond operator spin representation, we propose an approximate ground state
wavefunction which consistently covers all phases by means of a unitary
transformation.
The excitations can be efficiently described as independent bosons;
in the antiferromagnetic phase these reduce to the well-known spin waves, whereas
they describe gapped spin-1 excitations in the singlet phase.
We compute the spectra of these excitations as well as the
magnetizations throughout the whole phase diagram.

\end{abstract}

]

\section{Introduction}

Bilayer quantum magnets have attracted much interest in experiment and theory
in recent years, especially in the context of quantum Hall systems
and of quasi-two dimensional transition metal oxides.
On the experimental side,
quantum Hall systems \cite{sankar} are especially suitable for
investigating zero temperature quantum transitions between states with
different spin magnetizations.
In particular,
bilayer quantum Hall systems at filling fraction $\nu = 2$
~\cite{pellegrini,sawada,macdonald} can be tuned between
a fully polarized, ferromagnetic state and a spin singlet ground
state as function of the layer distance.
It is now well established \cite{pellegrini,pinczuk,sarma,demler}
that there is not a first-order ferromagnet--singlet transition,
but rather an intermediate phase with {\em canted} spin ordering
bounded by second-order transitions.
The spin degrees of freedom therefore appear to be well described
by a bilayer quantum spin model \cite{troyer}.
On the other hand, transition metal oxides like cuprates are known to
form two-dimensional structures where the low-energy spin dynamics is
well described by a Heisenberg model.
These materials consist either of a single plane or a stack of copper oxide planes
with intervening charge reservoir layers.
Prominent examples for bilayer compounds are $\text{YBa}_2\text{Cu}_3\text{O}_{7-x}$
and ${\rm Bi}_2 {\rm Sr}_2 {\rm CaCu}_2 {\rm O}_{8+\delta}$
which show a number of unusual properties \cite{mimo93}.
Note that in the high-temperature superconductors the interlayer coupling is
relatively weak compared to intralayer processes;
undoped bilayer systems like $\text{YBa}_2\text{Cu}_3\text{O}_6$ are not driven into a
singlet ground state.
However, the latter is also possible: recently the material BaCuSi$_2$O$_6$ containing
strongly coupled bilayers has been discovered and investigated \cite{sasago},
it shows a spin gap due to strong antiferromagnetic exchange interaction between
the two layers.

On the theoretical side, the bilayer Heisenberg magnet has attracted a lot of interest
\cite{troyer,sandvik,SaChuSa,weihong,kotov,matsushita,deng,mimo94,chub95,hida,gelfand,ng,elstner,zaanen,quiang,sandvik2}
because it is a simple model for studying the interplay of
long-range magnetic order and quantum disorder.
Quantum-critical behavior associated with such a magnetic instability
has been observed, {\em e.g.}, in the cuprate superconductors over
a wide range of doping levels and temperatures \cite{afscale}.

We start by describing the bilayer quantum Heisenberg model.
The system consists of two planes of nearest-neighbor $S=\frac{1}{2}$ Heisenberg
models with coupling constant $J_\parallel$ which can be either
antiferromagnetic ($J_\parallel > 0$) or ferromagnetic ($J_\parallel<0$).
The spins of corresponding sites of each layer are coupled
antiferromagnetically with a coupling constant $J_\perp > 0$.
The Hamiltonian reads
\begin{equation}
{\cal H} =
  J_{\perp}        \sum_i{\bf S}_{i1} \!\cdot\! {\bf S}_{i2}
+ J_{\parallel} \! \sum_{\langle ij\rangle m} {\bf S}_{i m} \!\cdot\! {\bf S}_{j m}
- {\bf B}\!\cdot\! \sum_{i m} {\bf S}_{i m}
\label{Hamiltonian}
\end{equation}
where ${\bf S}_{i m}$ are the electronic spin operators.
The index $i$ denotes {\em rungs} of the bilayer lattice, and $m=1,2$ labels the planes.
The summation $\left<ij\right>$ runs over pairs of nearest-neighbor rungs.
The external magnetic field ${\bf B}$ is homogeneous, the Zeeman interaction
factors $g_J \mu_B$ (gyromagnetic ratio and Bohr magneton) have been
absorbed in the definition of ${\bf B}$.

\begin{figure}
\epsfxsize=2.9in
\centerline{\epsffile{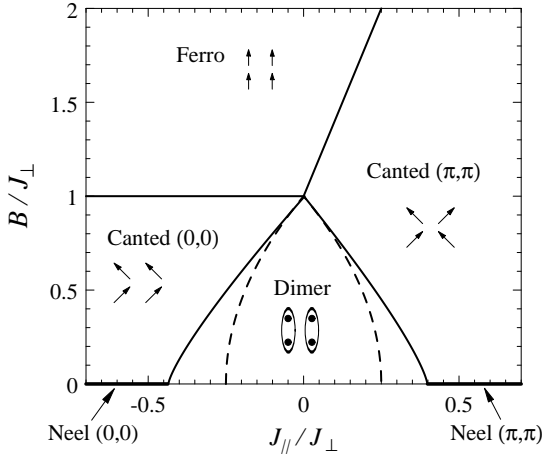}}
\caption{
Ground state phase diagram of $\cal H$ (\protect\ref{Hamiltonian})
as determined from series expansion studies, see
Refs~\protect\onlinecite{troyer,matsushita,gelfand}.
The arrows denote the spin orientations, the in-plane ordering wavevectors
are given in brackets.
The dashed lines are the boundaries of the singlet phase
obtained from the lowest-order boson approximation as described
in Sec~\protect\ref{sec:bosons}.
}
\label{fig_pd}
\end{figure}

In the following, we briefly discuss the main features of the ground-state phase
diagram of the model (\ref{Hamiltonian}), which has
been given in two recent papers \cite{troyer,matsushita} and
is shown in Fig.~\ref{fig_pd}.
In the absence of a magnetic field the bilayer Heisenberg antiferromagnet is known to
exhibit quantum phase transitions between
a disordered singlet phase and long-range ordered (LRO) phases.
In the limit of strong interplanar coupling, $J_{\perp} \gg |J_{\parallel}|$,
the system consists of weakly interacting rung singlets, and the ground state possesses
the full symmetry of the Hamiltonian.
The spin excitations are triplet modes with a minimum energy gap $\Delta$, there is no
magnetic LRO.
In the opposite case of large $|J_\parallel|/J_\perp$, the system possesses LRO
at $T=0$, the ordering wavevector is $(0,0,\pi)$ [$(\pi,\pi,\pi)$] for
ferromagnetic [antiferromagnetic] $J_\parallel$.
The SU(2) symmetry of $\cal H$ is broken, the low-lying excitations are doubly-degenerate
Goldstone spin waves.
The quantum transitions between the singlet and the two ordered phases
are of the O(3) universality class \cite{sandvik,SaChuSa,chub95,zaanen}
and occur at critical ratios $(J_\perp/J_\parallel)_{c1,2}$.
For the antiferromagnetic case,
quantum Monte Carlo calculations \cite{sandvik,SaChuSa,sandvik2}, series expansions
\cite{weihong}, and the diagrammatic Brueckner approach \cite{kotov}
yield an order-disorder transition point of
$(J_\parallel/J_\perp)_{c1} = 0.396$.
Bond-operator mean-field theory applied to the
bilayer Heisenberg AF \cite{matsushita,deng} gives a transition
point of $(J_\parallel/J_\perp)_{c1} = 0.435$.
Note that Schwinger boson mean-field theory \cite{mimo94} predicts a
value of $(J_\parallel/J_\perp)_{c1} \approx 0.22$, and also
self-consistent spin-wave theory \cite{chub95,hida},
which yields $(J_\parallel/J_\perp)_{c1} \approx 0.23$, fails to
reproduce the numerical results.
As Chubukov and Morr \cite{chub95} have pointed out
this discrepancy is due to the neglect of longitudinal
spin fluctuations in the conventional spin-wave approach.
For ferromagnetic $J_\parallel < 0$, the series expansion result is
$(J_\parallel/J_\perp)_{c2} = -0.435$ which agrees well with the
value from bond-operator mean-field theory.

For large enough external field, ${\bf B}=B{\bf e_z}$,
the exact ground state is the fully polarized ferromagnetic (FPF) state, {\em i.e.},
the state with all spins pointing in the direction of the applied field.
The first excited state is a single spin flip and its excitation energy can be determined exactly:
$\omega_{\bf k}=B-J_\perp-J_\parallel(2-\cos k_x-\cos k_y)$.
For ferromagnetic [antiferromagnetic] intra-plane coupling $J_\parallel$
the minimum of $\omega_{\bf k}$ is located
at in-plane wavevector ${\bf Q}=(0,0)$ [${\bf Q}=(\pi,\pi)$].
The stability boundary of the FPF state is given by the point where the
minimum excitation energy vanishes,
this yields the exact expression for the upper critical field,
$B_{c2}= J_\perp$ [$B_{c2} = 4J_\parallel +J_\perp$] for ferromagnetic [antiferromagnetic]
$J_\parallel$.
Single spin flips condense at this boundary, leading to canted spin ordering.
The transition is of second order and in the universality class of the dilute Bose gas
quantum-critical point with $z=2$ \cite{sarma,demler,troyer,matsushita,sase96}.
For intermediate magnetic fields a canted spin phase is established which
breaks the rotational symmetry of the Hamiltonian about the $z$ axis,
leading to one linear dispersing Goldstone mode corresponding to a rotation of
the order parameter in the $x$-$y$ plane.
Both the expectation value of the uniform magnetization in field direction
and the {\bf Q}-staggered magnetization perpendicular to the field are non-zero.

For small uniform fields, $B < \Delta$, the disordered spin singlet ground
state discussed above remains unchanged due to its finite excitation
gap $\Delta$.
The effect of the magnetic field is simply to split the degenerate
triplet excitations due to Zeeman coupling,
$\omega_{\bf k} \rightarrow \omega_{\bf k} - m B$, where $m=-1,0,1$ is
the $S_z$ quantum number.
The stability boundary of the singlet phase is determined by the vanishing
of the excitation gap;
the corresponding transition is in the same universality class as the boundary of
the fully polarized phase, also leading to the same canted phase with
ordering vector $\bf Q$.

The purpose of the present paper is to propose a transformation of the usual
bond operator basis which allows an efficient description of {\em all} phases
described above.
In the simplest approximation, the excitations can be described as independent
bosons.
In the limit of weak inter-plane coupling, the approach then reduces to linear
spin-wave theory, with the surplus of correctly describing the gapped
{\em longitudinal} amplitude mode which corresponds to inter-plane quantum
fluctuations and becomes important near the transition to a singlet state.
The paper is organized as follows:
In Sec~\ref{sec:trafo} we introduce the bond operator transformation which is employed to
describe the magnetic phases of the bilayer model.
Sec~\ref{sec:bosons} is devoted to the boson approximation which can be used
to diagonalize the Hamiltonian via a Bogoliubov transformation.
Results for ground state magnetizations and excitation spectra are
given in Sec~\ref{sec:results}.
A summary and discussion of possible improvements beyond
the independent boson approximation close the paper.


\section{Generalized bond operators}
\label{sec:trafo}

The four spin states per rung can be conveniently described via
a bond operator representation \cite{sachdev_1} of the two spins
${\bf S}_{i1}$ and ${\bf S}_{i2}$ of each rung $i$.
We introduce bosonic bond operators for creation
of a singlet and three triplet states out of the vacuum $|0\rangle$:
\begin{eqnarray}
s_i^{\dagger}\left|0\right>
&\!=\!& \frac{1}{\sqrt{2}} (c_{i1,\uparrow}^\dagger c_{i2,\downarrow}^\dagger \!-\!
                        c_{i1,\downarrow}^\dagger c_{i2,\uparrow}^\dagger)\left|0\right>
=   \frac{1}{\sqrt{2}}\left(\left|\uparrow\downarrow\right> \!-\!
                      \left|\downarrow\uparrow\right>\right)
\label{bond-op-repr} \\
t_{ix}^{\dagger}\left|0\right>
&\!=\!& \frac{-1}{\sqrt{2}}(c_{i1,\uparrow}^\dagger c_{i2,\uparrow}^\dagger\!-\!
                        c_{i1,\downarrow}^\dagger c_{i2,\downarrow}^\dagger)\left|0\right>
=  \frac{-1}{\sqrt{2}}\left(\left|\uparrow\uparrow\right> \!-\!
                      \left|\downarrow\downarrow\right>\right)
\nonumber \\
t_{iy}^{\dagger}\left|0\right>
&\!=\!& \frac{i}{\sqrt{2}}(c_{i1,\uparrow}^\dagger c_{i2,\uparrow}^\dagger\!+\!
                       c_{i1,\downarrow}^\dagger c_{i2,\downarrow}^\dagger)\left|0\right>
=  \frac{i}{\sqrt{2}}\left(\left|\uparrow\uparrow\right>\!+\!
                     \left|\downarrow\downarrow\right>\right)
\nonumber \\
t_{iz}^{\dagger}\left|0\right>
&\!=\!& \frac{1}{\sqrt{2}}(c_{i1,\uparrow}^\dagger c_{i2,\downarrow}^\dagger\!+\!
                       c_{i1,\downarrow}^\dagger c_{i2,\uparrow}^\dagger)\left|0\right>
=  \frac{1}{\sqrt{2}}\left(\left|\uparrow\downarrow\right>\!+\!
                     \left|\downarrow\uparrow\right>\right)
\nonumber
\end{eqnarray}
where $c_{im\sigma}^\dagger$ are creation operators for electrons at site $im$ with
spin $\sigma$.
In terms of the bond operators the spin operators can be expressed as
\begin{equation}
S_{i1,2}^{\alpha} = \frac{1}{2} ( \pm s_i^{\dagger}  t_{i\alpha}^{} \pm
t_{i\alpha}^{\dagger} s_i^{}  - i \epsilon_{\alpha\beta\gamma} t_{i\beta}^{\dagger}
t_{i\gamma})
\,.
\label{spin-rep}
\end{equation}
A local constraint of the form
\begin{equation}
s_i^{\dagger}s_i+\sum_{\alpha} t_{i\alpha}^{\dagger}t_{i\alpha} = 1 \label{constraint}
\end{equation}
has to be imposed on each rung to ensure that the physical states are either
singlets or triplets.
The ground state for vanishing intra-plane coupling, $J_{\parallel}=0$, is given by the product
state of singlet bonds on each rung, {\em i.e.},
\begin{equation}
\left| \phi_0\right> = \prod_i s_i^{\dagger}\left|0\right> \,.
\label{prod-singl}
\end{equation}
In this limit, the excitations are localized triplets with an energy gap $J_{\perp}$.

Substituting the bond operator representation of the spin operators (\ref{spin-rep})
into the original Heisenberg model we obtain the following Hamiltonian
\cite{gopalan,eder98,vojta}:
\begin{eqnarray}
{\cal H}_0 &=& -\frac{3}{4} J_\perp\sum_i s^{\dagger}_i s_i +
                \frac{1}{4} J_\perp\sum_{i\alpha} t^{\dagger}_{i\alpha} t_{i\alpha} \,, \nonumber \\
{\cal H}_1 &=& \frac{J_\parallel}{2}\sum_{\langle ij\rangle ,\alpha}
               ( t^{\dagger}_{i\alpha} s^{\dagger}_{j} t_{j\alpha} s_{i} + h.c.)  \nonumber \\
           &+& \frac{J_\parallel}{2}\sum_{\langle ij\rangle ,\alpha}
               ( t^{\dagger}_{i\alpha} t^{\dagger}_{j\alpha} s_{i} s_{j} + h.c. ) \nonumber \\
           &+& \frac{J_\parallel}{2}\sum_{\langle ij\rangle ,\alpha,\beta}
           (t^{\dagger}_{i\alpha} t^{\dagger}_{j\beta} t_{j\alpha} t_{i\beta}-t^{\dagger}_{i\alpha} t^{\dagger}_{j\alpha} t_{i\beta} t_{j\beta} )
               \nonumber \\
           &+& iB\sum_{i} ( t^{\dagger}_{ix} t_{iy} - t^{\dagger}_{iy} t_{ix} )
\,.
\label{orig-hamiltonian}
\end{eqnarray}

In the following the basic idea of describing the various magnetic phases
will be discussed.
Starting from the singlet phase, the magnetically ordered phases can be
described by different triplet boson condensates \cite{chub95,vojta,bruce}.
The FPF phase with polarization in $z$ direction requires a condensate
of $(t_x + i t_y)$ bosons;
a N\'{e}el state with staggered magnetization perpendicular to the field
corresponds to a condensate of $t_x$ or $t_y$ on top of the singlet
state.
Neglecting inter-rung fluctuations, an ansatz wavefunction can be written as
\begin{eqnarray}
|\tilde{\phi}_0\rangle
&\sim& \exp(i\mu\sum_i t^{\dagger}_{iy} t_{ix})\exp(\lambda\sum_i \text{e}^{i{\bf QR}_i}t^{\dagger}_{ix} s_i)
|\phi_0\rangle
\nonumber \\
&=& \prod_i \bigl(s^{\dagger}_i+\lambda \text{e}^{i{\bf QR}_i}(t^{\dagger}_{ix}+i\mu t^{\dagger}_{iy})\bigr)|0\rangle
\,.
\label{field-trafo}
\end{eqnarray}
The real parameters $\lambda$ and $\mu$ are the condensation amplitudes,
$\bf Q$ denotes the in-plane ordering wavevector.
This ansatz contains two subsequent unitary transformations of the singlet product state
$|\phi_0\rangle$ (\ref{prod-singl}):
the $\lambda$ term creates a spin-density-wave condensate with ordering vector $\bf Q$ and
quantization axis in $x$ direction [thus explicitely breaking the U(1) symmetry of the
$z$ axis rotation],
the $\mu$ term introduces canting and a non-zero $z$ magnetization into this state.
For $\mu=0$ and ${\bf Q}=(\pi,\pi)$ which corresponds to $J_\parallel > 0$ and $B=0$,
this wavefunction reduces to a state interpolating between the singlet and the N\'{e}el
state; it has been recently used to describe the dynamics of holes doped into
a bilayer system \cite{vojta}.
A finite external field $B$ larger than the spin gap will lead to a finite uniform
magnetization described by finite $\mu$;
for large $B$ the system is driven into a fully polarized state with
$\lambda \to \infty$ and $\mu = 1$.
The intervening canted phase will be characterized by finite, non-zero $\lambda$ and $0 < \mu < 1$.

A related product state has recently been employed for the description
of bilayer quantum Hall systems \cite{demler}, with the inclusion
of charge fluctuations and additional disorder, but without
inter-rung spin fluctuations.
We will show here that such spatial correlations can be efficiently
described in a simple harmonic theory.

For a proper description of fluctuations around the product state
$|\tilde \phi_0\rangle$ (\ref{field-trafo})
it is convenient to transform the basis states
on each rung.
We replace the basis operators $\{s_i,t_{i\alpha}\}$ by
\begin{eqnarray}
\tilde{s}^{\dagger}_i &=&\frac{1}{\sqrt{1+\lambda^2}}
  \left(s^{\dagger}_{i}+\frac{\lambda \text{e}^{i\mathbf{Q}\mathbf{R_i}}}{\sqrt{1+\mu^2}}
  (t^{\dagger}_{ix}+i\mu t^{\dagger}_{iy})\right) \,, \nonumber \\[6pt]
\tilde{t}^{\dagger}_{ix} &=&\frac{1}{\sqrt{1+\lambda^2}}\left(-\lambda \text{e}^{i\mathbf{Q}\mathbf{R_i}}
  s^{\dagger}_{i}+\frac{1}{\sqrt{1+\mu^2}}\big(t^{\dagger}_{ix}+i\mu t^{\dagger}_{iy}\big)\right) \,, \nonumber \\[6pt]
\tilde{t}^{\dagger}_{iy} &=&\frac{1}{\sqrt{1+\mu^2}}
  \left(t^{\dagger}_{iy}+i\mu t^{\dagger}_{ix}\right) \,, \nonumber \\[6pt]
\tilde{t}^{\dagger}_{iz} &=& t^{\dagger}_{iz}
\,.
\label{field-op-trafo}
\end{eqnarray}
The constraint (\ref{constraint}) takes the form
$\tilde{s}_i^{\dagger}\tilde{s}_i+\sum_{\alpha} \tilde{t}_{i\alpha}^{\dagger}\tilde{t}_{i\alpha} = 1$.
The new product state $\vert \tilde{\phi}_0\rangle$ can now be written as
\begin{equation}
| \tilde{\phi}_0\rangle = \prod_i \tilde{s}_i^{\dagger}\left|0\right>
\label{field-prod-state}
\end{equation}
which reduces to the singlet product state for $\lambda=0$.
Excitations out of this state are given by $\tilde{t}^{\dagger}_{i\alpha}\tilde{s}_i$.
At this point it is worth emphasizing that the use of a {\em single} condensate
(of ``mixed'' $\tilde{s}$ bosons) and the basis transformation to excitation operators
being {\em orthogonal} to $\tilde{s}$ is crucial for a correct description
of the ordered phases.

We can now insert the new basis operators into (\ref{orig-hamiltonian}) and
obtain the expression for the transformed Hamiltonian $\tilde{{\cal H}}$:
\begin{eqnarray}
\tilde{{\cal H}}_0 &=& \sum_{i\alpha} H(\alpha;\alpha) \; \tilde{t}_{i\alpha}^\dagger \tilde{t}_{i\alpha}+
               \sum_i H(s;s) \; \tilde{s}_{i}^\dagger \tilde{s}_i
\nonumber \\
               &+& \sum_i H(x;y) \; i(\tilde{t}_{ix}^\dagger \tilde{t}_{iy}-
                                   \tilde{t}_{iy}^\dagger \tilde{t}_{ix}) \,, \nonumber \\[6pt]
\tilde{{\cal H}_1}&=& \sum_i H(x;s) \; (\tilde{t}_{ix}^\dagger \tilde{s}_{i}+
                                      \tilde{s}_{i}^\dagger \tilde{t}_{ix} ) \nonumber \\[6pt]
             &+& \sum_i H(y;s) \; i(\tilde{t}_{iy}^\dagger \tilde{s}_{i}-
                                      \tilde{s}_{i}^\dagger\tilde{t}_{iy} ) \nonumber \\[6pt]
             &+& \sum_{\langle ij\rangle} H(ss;ss)\; \tilde{s}_i^\dagger\tilde{s}_j^\dagger\tilde{s}_i\tilde{s}_j \nonumber \\[6pt]
            &+& \sum_{\langle ij\rangle\alpha} H(\alpha\alpha;ss) \; \tilde{t}_{i\alpha}^\dagger\tilde{t}_{j\alpha}^\dagger
                                     \tilde{s}_i\tilde{s}_j+h.c. \nonumber \\[6pt]
            &+& \sum_{\langle ij\rangle} H(xy;ss) \; \tilde{t}_{ix}^\dagger\tilde{t}_{jy}^\dagger\tilde{s}_i\tilde{s}_j
             + h.c. + i \leftrightarrow j \nonumber \\[6pt]
            &+& \sum_{\langle ij\rangle\alpha} H(\alpha s;s\alpha) \; \tilde{t}_{i\alpha}^\dagger\tilde{t}_{j\alpha}
                                                         \tilde{s}_j^\dagger\tilde{s}_i+h.c.
\nonumber \\
             &+& \sum_{\langle ij\rangle} H(xs;sy) \; \tilde{t}_{ix}^\dagger\tilde{t}_{jy}\tilde{s}_j^\dagger\tilde{s}_i
             + h.c. + i \leftrightarrow j \nonumber \\[6pt]
            &+& \sum_{\langle ij\rangle} H(xs;xs) \; \tilde{t}_{ix}^\dagger\tilde{t}_{ix}\tilde{s}_j^\dagger\tilde{s}_j
             + i \leftrightarrow j
\nonumber \\
             &+& \sum_{\langle ij\rangle} H(ys;ys) \; \tilde{t}_{iy}^\dagger\tilde{t}_{iy}\tilde{s}_j^\dagger\tilde{s}_j
             + i \leftrightarrow j \nonumber \\[6pt]
            &+& \sum_{\langle ij\rangle} H(xs;ys) \; \tilde{t}_{ix}^\dagger\tilde{t}_{iy}\tilde{s}_j^\dagger\tilde{s}_j
             + i \leftrightarrow j \nonumber \\[6pt]
            &+& \sum_{\langle ij\rangle} H(xs;ss) \; \left( \tilde{t}_{ix}^\dagger\tilde{s}_{i}
                                               \tilde{s}_j^\dagger\tilde{s}_j +
                                               \tilde{s}_{j}^\dagger\tilde{t}_{ix}
                                               \tilde{s}_j^\dagger\tilde{s}_j\right) + i \leftrightarrow j \nonumber \\[6pt]
            &+& \sum_{\langle ij\rangle} H(ys;ss) \; \left( \tilde{t}_{iy}^\dagger\tilde{s}_{i}
                                               \tilde{s}_j^\dagger\tilde{s}_j +
                                               \tilde{s}_{j}^\dagger\tilde{t}_{iy}
                                               \tilde{s}_j^\dagger\tilde{s}_j\right) + i \leftrightarrow j \nonumber \\[6pt]
            &+& \tilde{{\cal H}}_{\text{rest}}
\,.
\label{field-hamiltonian}
\end{eqnarray}
The coefficients $H(\cdot,\cdot)$ depend on the basis parameters $\lambda$ and $\mu$ as
well as on $J_{\parallel}$, $J_{\perp}$, and $B$;
explicit expression are given in Appendix A.
$\tilde{{\cal H}}_{\text{rest}}$ contains more complicated terms with three and
four excitations.
In the case of $\mu=0$ the Hamiltonian reduces to the Hamiltonian derived for
the zero-field case \cite{vojta}.
It can be seen that $\tilde{{\cal H}}_1$ contains creation, hopping, and conversion terms
of the three types of excitations $\{\tilde{t}_{ix},\tilde{t}_{iy},\tilde{t}_{iz}\}$.
Contrary to the original Hamiltonian ${\cal H}_1$ where the triplet excitations can only occur in
pairs, here also single $\tilde{t}_{ix}$ and  $\tilde{t}_{iy}$ excitations can be created.
The effect of these terms is directly related
to the basis parameters $\lambda$ and $\mu$ and will be used to determine their values.
In contrast to the case of zero field where only pairs of the
same type of excitation are created out of $\vert \tilde{\phi}_0\rangle$,
for any finite $B$ also pairs of the structure
$\tilde{t}_{ix}^\dagger\tilde{t}_{jy}^\dagger\tilde{s}_i\tilde{s}_j$ are present.

The exact but lengthy representation (\ref{field-hamiltonian}) can now
be used for a variety of approximation schemes.
The most important point is to choose values for $\lambda$ and $\mu$ in such a way
that the product state $|\tilde{\phi}_0\rangle$ (\ref{field-prod-state})
is a reasonable starting point for approximations.
Fluctuations are described by the hard-core bosons
$\{\tilde{t}_{ix},\tilde{t}_{iy},\tilde{t}_{iz}\}$,
and a ``good'' choice for $|\tilde{\phi}_0\rangle$ should ensure that
the ground-state density of the bosons is small.
The lowest-order approximation corresponds to independent bosons
with a bilinear Hamiltonian,
it is formally given by ignoring both the hard-core constraint (\ref{constraint})
as well as $\tilde{\cal H}_{\rm rest}$,
and assuming a complete condensation of the generalized ``singlet'', {\em i.e.},
$\langle\tilde{s}\rangle = \tilde{s}= 1$.
This approximation will be discussed in the following sections,
it is shown to correspond to usual linear spin-wave theory in the
decoupled plane limit with $B=0$.
Possible improvements include (i) taking into account the site-averaged constraint
by introduction of a chemical potential $\lambda_0$, and treating
$\lambda_0$ and $\tilde s$ as variational parameters in the spirit of
bond-operator mean-field theory \cite{matsushita,deng},
(ii) a mean-field-like factorization of the higher-order boson interaction
terms
(iii) a diagrammatic treatment of the hard-core boson interaction in
the framework of Brueckner theory \cite{kotov}.
Independently, methods focussing on higher-order local excitations based
on cumulants \cite{vojta} or the coupled-cluster technique \cite{cc}
known from quantum chemistry may be applied
to (\ref{field-hamiltonian}).


\section{Independent Boson Approximation}
\label{sec:bosons}

In this and the following section, we will discuss the lowest-order boson
approximation for the Hamiltonian (\ref{field-hamiltonian}).
For this purpose we treat the excitations $\{\tilde{t}_{ix},\tilde{t}_{iy},\tilde{t}_{iz}\}$
as independent bosons, {\em i.e.}, neglect the constraint which restricts the Hilbert space
per rung.
The (generalized) singlet is assumed to fully condense, $\langle\tilde{s}\rangle = 1$.
It is known from the zero-field case \cite{kotov,vojta}
that this approximation is controlled
by the existence of a small parameter, namely the density of (generalized) triplet
excitations $\langle\tilde{t}_{i\alpha}^\dagger\tilde{t}_{i\alpha}\rangle$.
With this in mind, all terms of the Hamiltonian containing more than two
excitation operators $\tilde{t}_{i\alpha}$ will be neglected.

The parameters of the unitary transformation $\lambda$ and $\mu$ have not
yet been determined.
Within the approximation of independent bosons these parameters are chosen so that
the prefactors of the terms creating single $\tilde{t}_{i\alpha}$ bosons out of
$\vert \tilde{\phi}_0\rangle$ vanish.
The physical meaning of this is easily understood: terms creating single bosons would
change the condensate densities (and therefore alter the effective values of
$\lambda$ and $\mu$), however, our aim is to fully account for the boson
condensation by the transformed basis state $|\tilde\phi_0\rangle$.
One arrives at the following non-linear equations for the basis
parameters $\lambda$ and $\mu$:
\begin{eqnarray}
\lambda^2 &=& \frac{\big[\pm 4J_\parallel-J_\perp(1\!+\!\mu^2)+2B\mu\big]\big(1\!+\!\mu^2\big)}
              {\big[\pm 4J_\parallel+J_\perp(1\!+\!\mu^2)-2B\mu\big]
              \big(1\!+\!\mu^2\big)+8J_\parallel\mu^2} \label{l-AFM}
\\
0 &=& B(1\!-\!\mu^4)(1\!+\!\lambda^2) \mp 4J_\parallel\mu(1\!+\!\mu^2)-4J_\parallel\lambda^2\mu(1\!-\!\mu^2)
\nonumber
\end{eqnarray}
where upper (lower) sign refers to $J_\parallel>0$ ($J_\parallel<0$).
Notably, these equations also follow from a variational principle, {\em i.e.},
they are obtained by minimizing $\langle\tilde\phi_0|{\cal H}|\tilde\phi_0\rangle$
with respect to $\lambda$ and $\mu$.
This clarifies the meaning of $|\tilde\phi_0\rangle$ as best variational state
without inter-rung fluctuations.

The system of nonlinear equations (\ref{l-AFM}) allows for the computation of the
phase boundaries. The rotational invariant spin
singlet phase is characterized by the value $\lambda=0$.
Solutions for $\mu$ are then obtained for $B < B_{c1}$, where $B_{c1}$ is the stability
boundary of the singlet phase:
\begin{eqnarray}
B_{c1} &=& \sqrt{J_\perp-4J_\parallel} ~~ (J_\parallel > 0) \,,\nonumber \\
B_{c1} &=& \sqrt{J_\perp+4J_\parallel} ~~ (J_\parallel < 0) \,.
\label{lower-boundary}
\end{eqnarray}
It can be seen that the singlet phase exists for $|J_\parallel/J_\perp| < 0.25$.
At the critical ratio $(J_\parallel/J_\perp)_c=\pm 0.25$ the
character of the zero-field ground state changes
from spin singlet to long-range order.
As discussed in Ref~\onlinecite{zaanen}, this {\em quasi-classical} value
of $(J_\parallel/J_\perp)_c$ is necessarily smaller than the exact one
(quantum fluctuations stabilize the inter-plane singlet phase);
as noted in the introduction the critical values obtained by large-scale numerics are
$(J_\parallel/J_\perp)_{c1}=0.396$ and $(J_\parallel/J_\perp)_{c2}=-0.435$.
The inclusion of higher order terms beyond the present linear ``spin-wave''
approximation leads to a better agreement with the numerical values
for the critical coupling ratio \cite{vojta}.

For values of the field with $B_{c1} < B < B_{c2}$ the
ground state is a canted phase with $0 < \mu < 1$.
The phase boundary to the ferromagnetic phase is obtained when the
limit $\lambda \rightarrow \infty$ and $\mu \rightarrow 1$
is taken.
We arrive at an upper critical field of
\begin{eqnarray}
B_{c2} &=& J_\perp+4J_\parallel ~~ (J_\parallel > 0) \,,\nonumber \\
B_{c2} &=& J_\perp ~~~~~~~~~~\,(J_\parallel < 0)
\label{upper-boundary}
\end{eqnarray}
which are the exact results.

We can now proceed with analyzing the fluctuations around $|\tilde\phi_0\rangle$.
With the approximations described above, we arrive at a bilinear Hamiltonian:
\begin{eqnarray}
\tilde{{\cal H}} &=& \sum_{{\bf k}\alpha} A_{{\bf k}\alpha}\tilde{t}^\dagger_{{\bf k}\alpha}\tilde{t}_{{\bf k}\alpha}
   + \frac{1}{2}\sum_{{\bf k}\alpha} B_{{\bf k}\alpha}\left(
    \tilde{t}^\dagger_{{\bf k}\alpha}\tilde{t}^\dagger_{{\bf -k}\alpha} +
    \tilde{t}_{{\bf k}\alpha}\tilde{t}_{{\bf -k}\alpha}\right)
\nonumber \\[6pt]
  &+& \sum_{\bf k} C_{\bf k}\left(\tilde{t}^\dagger_{{\bf k}x}\tilde{t}_{{\bf k}y}-
                                            \tilde{t}^\dagger_{{\bf k}y}\tilde{t}_{{\bf k}x} \right)
\nonumber \\
  &+& \sum_{\bf k} D_{\bf k}\left(\tilde{t}^\dagger_{{\bf k}x}\tilde{t}^\dagger_{{\bf -k}y}-
                                            \tilde{t}_{{\bf k}x}\tilde{t}_{{\bf -k}y} \right) \,.
\label{field-fourier-hamiltonian}
\end{eqnarray}
Here, $\tilde{t}_{{\bf k}\alpha}$ are the modified basis operators which have been Fourier
transformed with respect to the in-plane momentum $\bf k$.
The expressions for the coefficients $A_{\bf k}, \dots, D_{\bf k}$
can be found in Appendix B.
The Hamiltonian (\ref{field-fourier-hamiltonian}) can be easily diagonalized
with a Bogoliubov transformation,
leading to new generalized excitations denoted by
$\tau_{{\bf k}\alpha}$, $\alpha=x,y,z$.
The ground state $|\psi_0\rangle$ is simply the vacuum of the
bosons $\tau_{{\bf k}\alpha}$.
The excitation energies of these bosons are given by
\begin{eqnarray}
\Omega^2_{{\bf k}x,y} &=& \frac{1}{2}\left(\omega^2_{{\bf k}x}+\omega^2_{{\bf k}y} - 2C^2_{\bf k}+2D^2_{\bf k}\right) \nonumber \\
    &\pm& \biggl\{ \; \frac{1}{4}\left(\omega^2_{{\bf k}x}+\omega^2_{{\bf k}y} - 2C^2_{\bf k}+2D^2_{\bf k}\right)^2
    \nonumber \\
    &\phantom{{=}\pm}&
    - \omega^2_{{\bf k}x}\omega^2_{{\bf k}y} - \left(C^2_{\bf k}-D^2_{\bf k}\right)^2 \nonumber \\
    &\phantom{{=}\pm}& + 2\left(\left(A_{{\bf k}x}A_{{\bf k}y}-B_{{\bf k}x}B_{{\bf k}y}\right)
       \left(C^2_{\bf k}+D^2_{\bf k}\right) \right .
\nonumber \\
    &\phantom{{=}\pm}& \left .- 2\left(A_{{\bf k}x}B_{{\bf k}y}-A_{{\bf k}y}B_{{\bf k}x}\right)C_{\bf k}D_{\bf k}\right) \;
       \biggr\} ^\frac{1}{2} \,,
\nonumber \\
\Omega^2_{{\bf k}z} &=& A^2_{{\bf k}z}-B^2_{{\bf k}z} \,,
\label{dispers}
\end{eqnarray}
the terms $\omega^2_{{\bf k}x} := A^2_{{\bf k}x}-B^2_{{\bf k}x}$ and
$\omega^2_{{\bf k}y} := A^2_{{\bf k}y}-B^2_{{\bf k}y}$ are simply the eigenenergies
of the zero field limit.
In zero field, we find either three degenerate gapped modes in the singlet phase,
or two degenerate transverse acoustic spin-wave modes and a gapped spin-amplitude
mode in the long-range ordered phases.
A finite external field lifts these degeneracies; the canted phase supports
a single Goldstone mode due to the broken U(1) symmetry.
Explicit results for the dispersion relations are shown in the next section.

Finally the ground state energy per site is given
by the expression
\begin{equation}
E_g = \frac{\langle \psi_0|\tilde{\cal H}|\psi_0\rangle}{2N} =
E_0 + {1 \over 2N} \sum_{{\bf k}\alpha} (\Omega_{{\bf k}\alpha}-A_{{\bf k}\alpha}) \label{gs-energy}
\end{equation}
where $E_0$ is the energy per site of the product state $\vert \tilde{\phi}_0\rangle$,
$E_0 = \langle\tilde\phi_0|{\cal H}|\tilde\phi_0\rangle / 2N$.

The described harmonic approximation preserves SU(2) symmetry in
the singlet phase; in the ordered phases it
can be considered as linear spin-wave theory
for the bilayer problem with inclusion of {\em longitudinal} spin
fluctuations \cite{chub95,bruce,affleck} --
these are crucial for describing the magnetic
properties near the boundary to the singlet phase.
For zero field and vanishing inter-plane coupling, $B=J_\perp=0$,
the results of the present approach are equivalent to the ones of linear
spin-wave theory.
For instance, in the antiferromagnetic case, the magnetization takes
60 \% of its classical value.
There, the gapped amplitude mode is dispersionless and does not
influence the ground state
properties, and its spectral weight in neutron scattering
vanishes.


\section{Ground state properties and spin excitations}
\label{sec:results}

In this section we present the results obtained from the independent boson approximation.
We discuss the staggered and uniform magnetizations and the dispersion relations as
functions of the external magnetic field in order to characterize the various magnetic phases.
For this purpose we consider
the system for small intra-plane coupling $|J_\parallel|$, {\em i.e.},
in the singlet phase, as well as for weakly coupled planes (small $J_\perp$).

\begin{figure}
\epsfxsize=3in
\centerline{\epsffile{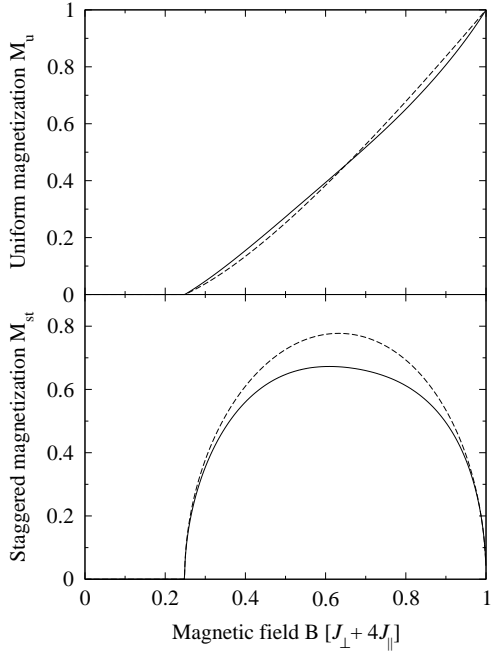}}
\caption{
Uniform (top) and staggered magnetization (bottom) as function of the external
magnetic field $B/B_{c2}$.
The ratio of the coupling constants $J_\parallel/J_\perp = 0.2$ which places
the zero-field system into the interlayer singlet phase.
Solid:  present approximation (independent bosons).
Dashed: product wavefunction $|\tilde{\phi}_0\rangle$ only.
}
\label{fig_mag1}
\end{figure}

\begin{figure}[t]
\epsfxsize=3in
\centerline{\epsffile{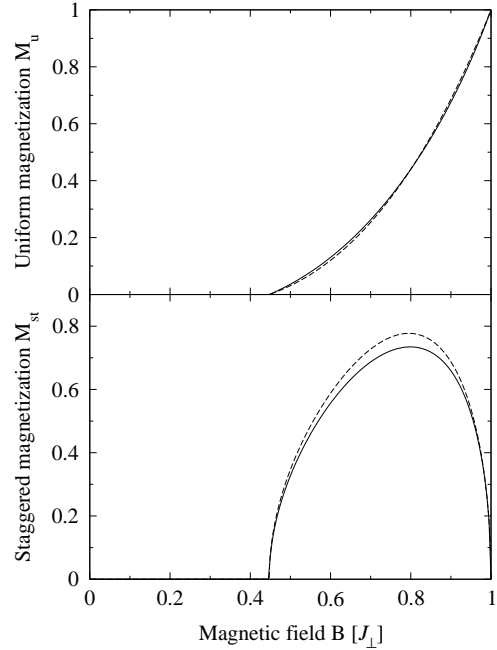}}
\caption{
As Fig.~\protect\ref{fig_mag1}, but in the singlet phase with strongly coupled
{\em ferromagnetic} planes with $J_\parallel/J_\perp = -0.2$.
}
\label{fig_mag2}
\end{figure}

\begin{figure}
\epsfxsize=3in
\centerline{\epsffile{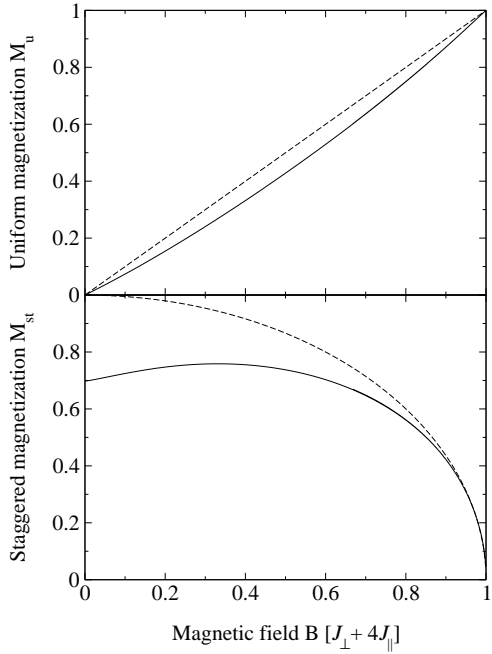}}
\caption{
As Fig.~\protect\ref{fig_mag1}, but for weakly coupled
antiferromagnetic planes with $J_\parallel/J_\perp = 10$.
}
\label{fig_mag3}
\end{figure}

\begin{figure}
\epsfxsize=3in
\centerline{\epsffile{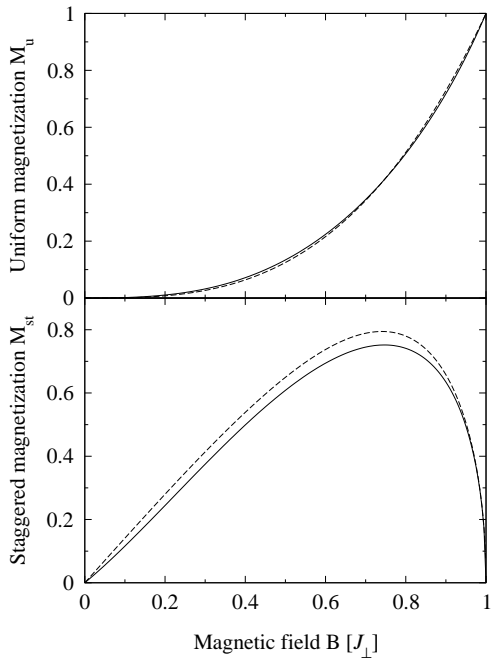}}
\caption{
As Fig.~\protect\ref{fig_mag1}, now at the zero-field critical point
between the singlet and ``ferromagnetic'' ${\bf Q}=(0,0)$ N\'{e}el
phase, $J_\parallel/J_\perp = -0.25$ (see Fig.~\protect\ref{fig_pd}).
}
\label{fig_mag4}
\end{figure}

\subsection{Magnetization}

We start with the uniform and staggered magnetizations which are obtained
as expection values of the corresponding spin operators
\begin{eqnarray}
M_u &=& \frac{1}{N} \sum_i \langle S_{i1}^z + S_{i2}^z \rangle \,,
\nonumber\\
M_{st} &=& \frac{1}{N} \sum_i \langle S_{i1}^x - S_{i2}^x \rangle \text{e}^{i{\bf QR}_i}
\end{eqnarray}
where ${\bf Q}=(0,0)$ [${\bf Q}=(\pi,\pi)$] for ferromagnetic [antiferromagnetic]
intra-plane coupling $J_\parallel$ as above.
The staggered magnetization in the zero-field case has been calculated earlier
(see Fig~2 of Ref~\onlinecite{vojta});
it shows a good overall agreement with the series expansion data of
Ref~\onlinecite{weihong} if plotted as function of
$(J_\parallel/J_\perp) / (J_\parallel/J_\perp)_c$ --
the critical coupling ratio itself is underestimated in the independent boson
approach as discussed above.

Turning to the finite-$B$ case, we show uniform and staggered magnetization in
Figs~\ref{fig_mag1}--\ref{fig_mag4}.
Both are zero in the singlet phase, $B<B_{c1}$.
Varying the magnetic field from the lower critical value $B_{c1}$ to the upper critical field $B_{c2}$
the uniform magnetization increases from $0$ and saturates at $B_{c2}$,
whereas the staggered magnetization is non-zero only in the canted phase,
$B_{c1} < B < B_{c2}$, passing a maximum in between
(Figs~\ref{fig_mag1}, \ref{fig_mag2}).
Near the critical fields for $B\gtrsim B_{c1}$ or $B\lesssim B_{c2}$, we have
$M_u \sim |B-B_{c1,2}|$ and $M_{st} \sim |B-B_{c1,2}|^{1/2}$.

For small $J_\perp$, {\em i.e.}, in the zero-field ordered regime,
the staggered magnetization $M_{st}$ is finite at $B=0$ but is suppressed if $B$ is
increased up to the critical field $B_{c2}$, see Fig~\ref{fig_mag3}.
Interestingly, $M_{st}$ increases (!) at small fields -- here the suppression of quantum
fluctuations due to the field ({\em e.g.} one Goldstone mode becomes gapped) has
a stronger effect than the change in the canting angle.
The uniform magnetization again increases continuously with $B$ until it saturates at
$B_{c2}$.
For illustration, we also show data for $J_\parallel/J_\perp=-0.25$ which places
the model at the zero-field critical point between the singlet and ${\bf Q}=(0,0)$
N\'{e}el phases. Here, $M_u$ increases with $B^2$ whereas $M_{st}$ raises linearly
with the applied field, in agreement with the results of Ref~\cite{sarma}.

In Figs~\ref{fig_mag1}--\ref{fig_mag4}, we have also shown the magnetization values
calculated with the product state $|\tilde{\phi}_0\rangle$ only (dashed lines) --
this is equivalent to neglecting intra-plane (inter-rung) fluctuations
altogether.
It can be seen that the effect of these spatial quantum fluctuations is largest
for antiferromagnetic $J_\parallel$ and large staggered magnetization, this
includes the case of the single-layer antiferromagnet where quantum fluctuations
are known to reduce $M_{st}$ by 40\%.
In contrast, in both the ferromagnetic regime (where quantum fluctuations are weak in
general) and near the spin singlet phase (where {\em inter-plane} quantum fluctuations
dominate), the effect of intra-plane fluctuations is weaker.

All phase transitions found are of second order, as expected,
except for the special point $J_\parallel=0$, $B=J_\perp$, where
the singlet--FPF transition is a simple level crossing for independent
rungs.
Due to the mean-field character of the present approximation, all
critical exponents have mean-field values.
(It is easily seen that the present harmonic approximation
becomes exact in the limit of large in-plane coordination number,
see also Ref~\onlinecite{zaanen}.)
For the actual two-dimensional model, these mean-field exponents
are incorrect for the zero-field transitions which have $z=1$
and therefore obey the non-trivial exponents of a three-dimensional
classical Heisenberg model,
but they apply (ignoring logarithmic corrections) to the field-induced
transitions from and to the canted phase which are at the
upper-critical dimension ($z=2$) \cite{sarma,demler,troyer,sase96}.

We note that the phase boundaries obtained in Sec~\ref{sec:bosons}
using the product state $|\tilde{\phi}_0\rangle$ are {\em not} changed
by the inclusion of inter-rung fluctuations at the lowest boson
level employed here.
It is, however, clear that boson interactions change this
picture \cite{kotov}.

\begin{figure}
\epsfxsize=3in
\centerline{\epsffile{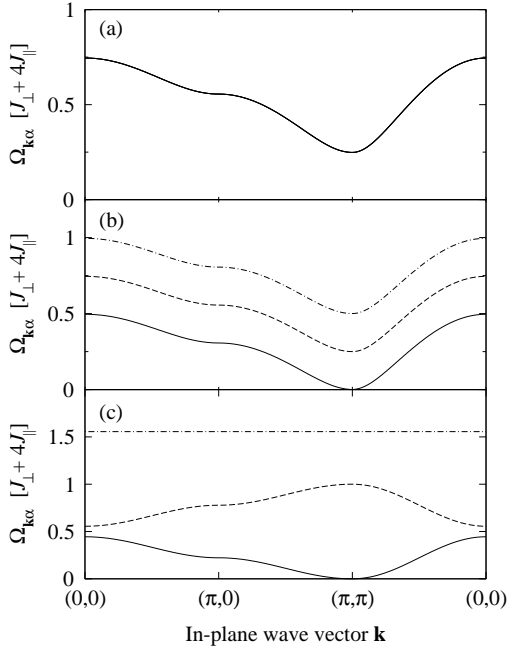}}
\caption{
Dispersion relations of the magnetic excitations for
strongly coupled antiferromagnetic planes, $J_\parallel/J_\perp = 0.2$.
The three panels show the three modes for different
magnetic fields $\tilde{B} = B/B_{c2} = B/(4J_\parallel+J_\perp)$,
(a) $\tilde{B}=0$, (b) $\tilde{B}=0.44$ (here $B=B_{c1}$),
(c) $\tilde{B}=1$ ($B=B_{c2})$.
}
\label{fig_disp1}
\end{figure}

\begin{figure}
\epsfxsize=3in
\centerline{\epsffile{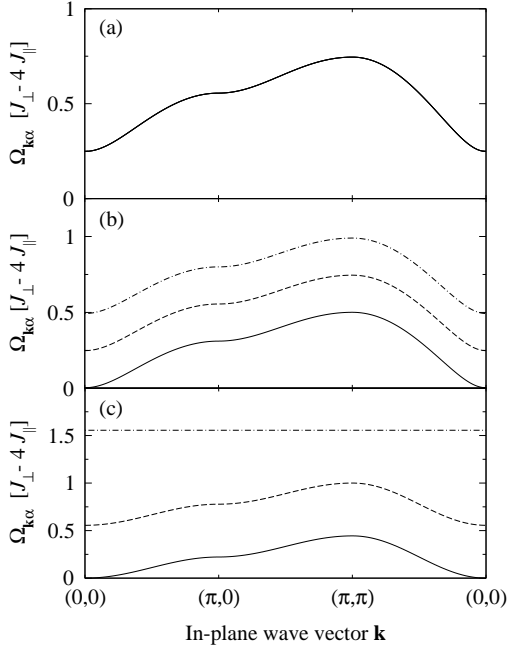}}
\caption{
As in Fig.~\protect\ref{fig_disp1}, but for
ferromagnetic in-plane coupling, $J_\parallel/J_\perp = -0.2$.
Note that the magnetic field values in units of $B_{c2}$ are again
$\tilde{B}=0$, 0.44, and 1, but here $B_{c2}=J_\perp$, {\em i.e.},
$\tilde{B}=B/J_\perp$.
}
\label{fig_disp2}
\end{figure}

\begin{figure}
\epsfxsize=3in
\centerline{\epsffile{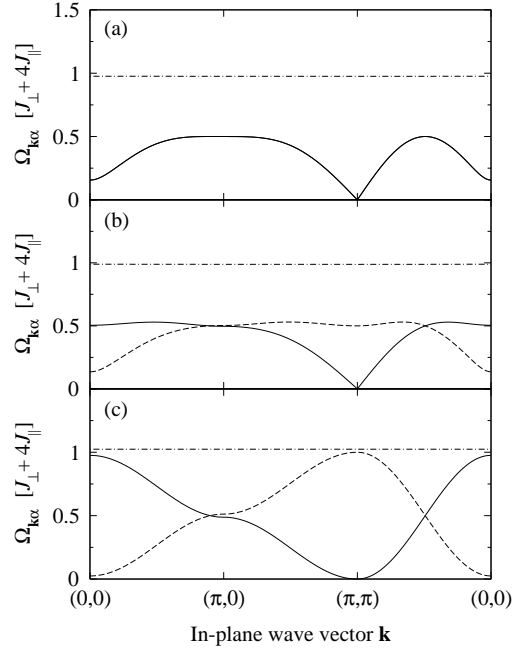}}
\caption{
As in Fig.~\protect\ref{fig_disp1}, here for
weakly coupled antiferromagnetic planes, $J_\parallel/J_\perp = 10$.
}
\label{fig_disp3}
\end{figure}

\begin{figure}
\epsfxsize=3in
\centerline{\epsffile{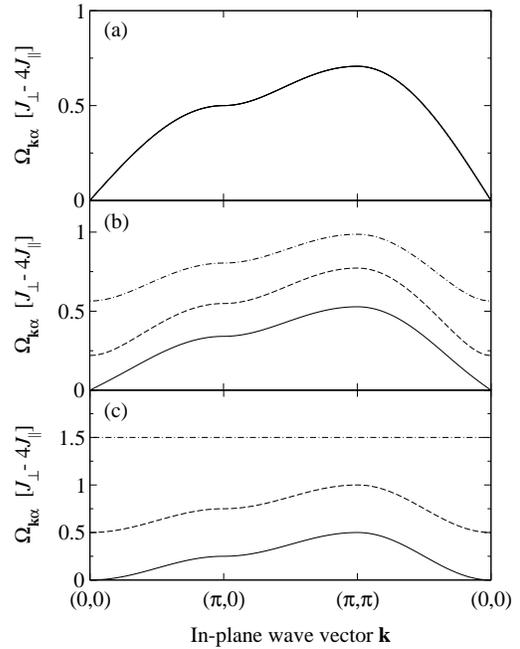}}
\caption{
As in Fig.~\protect\ref{fig_disp2}, finally for
the coupling constant ratio $J_\parallel/J_\perp = -0.25$,
corresponding to the zero-field critical point on the
ferromagnetic side of the phase diagram (Fig.~\protect\ref{fig_pd}).
}
\label{fig_disp4}
\end{figure}

\subsection{Magnetic excitations}

We continue with the discussion of the dispersion relations
of the magnetic excitations, plotted in Figs~\ref{fig_disp1}--\ref{fig_disp4},
for a number of values of the external field.
In Fig~\ref{fig_disp1} we show data for strongly coupled planes and antiferromagnetic
$J_\parallel$.
In the zero-field limit (Fig~\ref{fig_disp1}a) the excitations are threefold
degenerate, gapped, and
have a dispersion minimum at ${\bf Q}=(\pi,\pi)$.
At finite $B$, the degeneracy is lifted due to the Zeeman coupling.
At the lower critical field, $B=B_{c1}$ (Fig~\ref{fig_disp1}b),
the excitation gap closes, and there is a single critical mode
with quadratic dispersion.
In the canted phase, $B_{c1} < B < B_{c2}$ (not shown),
there is a single Goldstone mode with a linear dispersion
around ${\bf Q}=(\pi,\pi)$, corresponding to the broken $U(1)$ symmetry.
At the phase boundary to the ferromagnetic (FPF) phase,
$B=B_{c2}$ (Fig~\ref{fig_disp1}c), the lowest mode becomes
again critical with a quadratic dispersion (note $z=2$ at both field-induced
transitions).
As the magnetic field is further increased, a gap opens and
the FPF phase is stabilized.
In the FPF phase, one of the excitations is dispersionless,
corresponding to a longitudinal local fluctuation which turns both spins on a
rung down.
In the case of ferromagnetic in-plane coupling, $J_\parallel<0$, the
picture is similar, but the dispersion minimum is now found at ${\bf Q}=(0,0)$,
see Fig~\ref{fig_disp2}.

The results for weakly coupled antiferromagnetic planes are plotted in
Fig~\ref{fig_disp3}.
We start with a discussion of the zero field case, shown in Fig~\ref{fig_disp3}a.
Contrary to the case of strong inter-plane coupling there is now a
nearly dispersionless excitation -- this is the before-mentioned gapped (longitudinal)
spin-amplitude mode.
It corresponds to flipping both spins on a single rung and has a quantum number $S^z=0$,
consequently, it remains unchanged (to linear order) when the magnetic field
is turned on.
The two other modes are degenerate Goldstone bosons in the zero-field limit
with a linear dispersion around the ordering wavevector ${\bf Q}=(\pi,\pi)$.
For decoupled planes, $J_\perp\to 0$, momenta $(0,0)$ and $(\pi,\pi)$ are
degenerate, and for small $J_\perp$ the gap at $(0,0)$ is proportional to
$J_\perp$.
With increasing $J_\perp$, the spin-amplitude mode acquires a dispersion, until
it merges with the acoustic spin-wave modes at the zero-field critical point, leading
to a degenerate triplet of critical modes with linear dispersion (not shown).
Turning to the case of a finite uniform field $B$, we note that arbitrarily small $B$
establishes canted ordering here.
For $0 < B < B_{c2}$ (Fig~\ref{fig_disp3}b), the mode degeneracy is again lifted and
leaves a single gapless Goldstone excitation with linear dispersion.
At $B=B_{c2}$ (Fig~\ref{fig_disp3}c), this mode becomes critical with quadratic
dispersion as discussed above.

Finally, Fig~\ref{fig_disp4} shows dispersions for the coupling ratio corresponding
to the zero-field ferromagnetic critical point, $J_\parallel/J_\perp = -0.25$.
For $B=0$ (Fig~\ref{fig_disp4}a), we have a triplet of degenerate critical modes
with minimum (condensation) wavevector ${\bf Q}=(0,0)$ and a linear dispersion
reflecting $z=1$.
With application of a finite field, we enter directly the canted phase with a single
Goldstone mode, so for Figs~\ref{fig_disp4}b,c the above discussion applies.

At this point a remark about the physics of longitudinal magnon mode is in order.
As mentioned, this mode is not included in conventional spin-wave theory
for the antiferromagnet. The reason is that spin-wave theory has the
character of a $1/S$ expansion, and longitudinal fluctuations are suppressed in
the $S\to\infty$ limit.
This implies that any theory based on a (finite-order) large-$S$ expansion
can never treat longitudinal and transverse spin fluctuations on equal
footing~\cite{chub95},
which is, however, crucial near the critical point in the bilayer model where
longitudinal and transverse fluctuations become indistinguishable.
In contrast, our approach preserves the quantum nature of the $S=1/2$ spins,
but has the spirit of a large-$z$ theory, where $z$ is the
in-plane coordination number.

We close this section with some general remarks.
The lowest-order boson approximation neglects interactions between the
modes, therefore all excitations including the amplitude mode
are undamped, and no decay into the two-particle continuum is
possible.
The description of such damping processes requires the inclusion of
fourth-order boson terms, and is deferred to future investigation.

The low-energy properties of the spin excitations obtained in the present work
are in agreement
with hydrodynamical considerations.
The spin-wave velocity $c$ in an ordered phase,
defined as $\Omega_{\bf q} = c\,|{\bf q}-{\bf Q}|$ near
the ordering wavevector $\bf Q$, has to fulfill the relation
$c^2 = \rho_s / \chi_\perp$ where $\rho_s$ is the spin stiffness and
$\chi_\perp$ denotes the uniform susceptibility for a field perpendicular
to the ordering direction (here $x$).
At the zero-field transitions, $J_\parallel/J_\perp = (J_\parallel/J_\perp)_{c1,2}$,
both $\rho_s$ and $\chi_\perp$ vanish whereas $c$ remains non-critical
as expected for an O(3) transition.
This behavior is correctly reproduced in our calculations (Figs~\ref{fig_mag4} and
\ref{fig_disp4}), note $\chi_\perp = {\rm d} M_u / {\rm d} B$.
(In contrast, at the field-induced transition points the spin-wave velocity
$c$ vanishes, leading to a quadratically dispersing critical mode at $B=B_{c1,2}$.)

The special properties of the multicritical points at
$B=0$, $J_\parallel/J_\perp = (J_\parallel/J_\perp)_{c1,2}$ have been discussed
in Ref~\cite{troyer}.
One consequence of universal scaling arguments
is that the energy gap of the $B=0$ amplitude mode
in the ordered phase close to the transition
is given by the spin stiffness $\rho_s$; this result is special
to $d=2$ where $\rho_s$ has the dimension of energy.
We also note that a well-defined amplitude mode requires
that the low-energy theory for the zero-field ordering transition
is below its upper-critical dimension
(here $d_{\rm c,up}=3$) so that the interaction term of the effective
$\phi^4$ theory is relevant in the renormalization group
sense.
Our mean-field theory gives an amplitude mode independent of the
spatial dimension $d$, but
from the above it is clear that even the low-energy part of this mode
will be overdamped from boson interactions in higher dimensions, $d \geq 3$.

The results obtained with the present method for vanishing inter-plane
coupling, $J_{\perp}=0$, are consistent with recent calculations
using linear spin-wave theory \cite{zhitomirsky}
for a single-plane antiferromagnet.


\section{Conclusion}
\label{sec:concl}

In this paper, we have proposed a unitary transformation of bond-boson operators
appropriate for the description of various magnetic phases, especially
with canted spin order.
The resulting boson Hamiltonian has been obtained exactly; it allows for a number of
approximations.
We have applied this method to the bilayer Heisenberg magnet in a
uniform external field.
Using  the lowest-order approximation of free bosons we obtained analytic expressions
for the phase boundaries, the quasiparticle excitations and both uniform and staggered
magnetization.
The harmonic approximation is controlled by the smallness of the density
of generalized bosonic triplet excitations, this parameter
has been found to be smaller than 0.15 throughout the whole
phase diagram.
In the limit of vanishing inter-plane coupling, the boson approximation
reduces to linear spin-wave theory;
near the transition to the singlet phase it has the advantage
of incorporating {\em longitudinal} spin fluctuations.
The phase diagram (Fig~\ref{fig_pd}) is adequately described;
the locations of the zero-field transitions are $(J_\parallel/J_\perp)_{c1,2} = \pm 0.25$,
these values are smaller in magnitude than the ones obtained by more accurate numerical
and analytical methods
[$(J_\parallel/J_\perp)_{c1}=0.396$ and $(J_\parallel/J_\perp)_{c2}=-0.435$].
These deviations arise from the neglect of boson interactions which stabilize the
singlet phase.
%
The approach presented here preserves all general properties expected from
hydrodynamics, {\em i.e.}, it gives the correct number and low-energy dispersion
of Goldstone modes in the symmetry-broken phases.

The present approximation can be systematically improved as discussed in
Sec~\ref{sec:trafo}.
Two promising possibilities are either the inclusion of fourth-order boson
terms as in non-linear spin-wave theory
or the treatment of the hard-core boson constraint by introduction
of an (infinite) on-site repulsion which is treated diagrammatically
using Brueckner theory \cite{kotov}.
(The latter approach yields very accurate results for the zero-field
bilayer antiferromagnet.)
We expect that the inclusion of interactions will induce a damping
of the high-energy spin excitation modes, but will not change the
low-energy properties presented in Sec~\ref{sec:results} which are
protected by symmetry considerations and hydrodynamics.
Once interactions are included, a detailed study of the dynamic
structure factor as measured in inelastic neutron scattering,
including the spectral weights of the excitation modes described
here, would be interesting.

We close by mentioning possible applications of the present approach.
As in Ref \onlinecite{vojta}, it can be used for studying a hole- or
electron-doped system; in this case the ground state wavefunction derived here
can be used as a background state for the carrier motion.
Other interesting prospectives include the study of disorder
\cite{demler} and
the application to {\em frustrated} systems in the presence of
an external field.
A nice example here is the recently discovered triangular bilayer $S=1$ system
Ba$_3$Mn$_2$O$_8$ which shows a spin gap, magnetization plateaus, and
interesting frustration effects \cite{uchida01}.

\acknowledgments
{
The authors acknowledge useful discussions with C.~K\"uhnert, K.~Meyer,
and S.~Sachdev as well as financial support by the DFG (SFB 463 and 484).
}

\appendix
\section{Hamiltonian in the modified bond boson basis}

In the following, we list the first terms appearing in the transformed
bilayer Hamiltonian $\tilde{\cal H}$ (\ref{field-hamiltonian}) for
antiferromagnetic $J_\parallel$.
\begin{eqnarray*}
H(s;s) &=& \left(J_{\perp}+2B\frac{\mu}{1+\mu^2}\right)\frac{\lambda^2}{1+\lambda^2} \,, \\
H(x;x) &=& \left(J_{\perp}+2B\frac{\mu}{1+\mu^2}\right)\frac{1}{1+\lambda^2} \,, \\[12pt]
H(y;y) &=& J_{\perp}+2B\frac{\mu}{1+\mu^2} ~, ~~~H(z;z) = J_{\perp}  \,, \\[12pt]
H(x;s) &=& \left(J_{\perp}-2B\frac{\mu}{1+\mu^2}\right)
           \frac{\lambda\text{e}^{i\mathbf{Q}\mathbf{R}_i}}{1+\lambda^2} \,, \\[12pt]
H(y;s) &=& B \frac{1-\mu^2}{1+\mu^2}
        \frac{\lambda\text{e}^{i\mathbf{Q}\mathbf{R}_i}}{\sqrt{1+\lambda^2}} \,, \\[12pt]
H(x;y) &=& \frac{-B}{\sqrt{1+\lambda^2}} \frac{1-\mu^2}{1+\mu^2} \,,
\end{eqnarray*}
\begin{eqnarray*}
H(xx;ss) &=& \frac{J_{\parallel}}{2}
                \frac{\left(1\!+\!\mu^2\right) \left[ \left(1\!-\!\lambda^2\right)^2+\mu^2 \!
                      \left(1\!+\!\lambda^2\right)^2 \right] \!-\! 4\lambda^2\mu^2}
                     {\left(1+\lambda^2\right)^2\left(1+\mu^2\right)^2}  \,, \\[12pt]
H(yy;ss) &=& \frac{J_{\parallel}}{2}\;\;
                \frac{\left(1-\mu^2\right)\left[\left(1+\mu^2\right)+\lambda^2\left(1-\mu^2\right)\right]}
                     {\left(1+\lambda^2\right)\left(1+\mu^2\right)^2}  \,, \\[12pt]
H(zz;ss) &=& \frac{J_{\parallel}}{2}\;\;
                \frac{\left(1+\mu^2\right)+\lambda^2\left(1-\mu^2\right)}
                     {\left(1+\lambda^2\right)\left(1+\mu^2\right)} \,.
\end{eqnarray*}
These and the remaining terms can be obtained with the help of {\em Mathematica} by inserting
the transformation (\ref{field-op-trafo}) into ${\cal H}$ (\ref{orig-hamiltonian}).

\section{Bilinear Hamiltonian}

We list here the coefficients of the bilinear boson Hamiltonian
(\ref{field-fourier-hamiltonian}) which enter the expressions for
the dispersions given in (\ref{dispers}).
For shortness, we restrict ourselves to the antiferromagnetic case,
$J_\parallel>0$, and
use the abbreviation $\gamma_{\bf k} = \frac{1}{2}\left(\cos k_x + \cos k_y\right)$.
\begin{eqnarray*}
A_{{\bf k}x}
&=& \left(J_{\perp}-\frac{2B\mu}{1+\mu^2}\right) \frac{1-\lambda^2}{1+\lambda^2} +
8J_\parallel\frac{\lambda^2\left(2+3\mu^2-\lambda^2\mu^2\right)}{\left(1+\lambda^2\right)^2\left(1+\mu^2\right)^2}\\
\\[6pt]
&+&2J_{\parallel}\gamma_{\bf k}\frac{\left(1-\lambda^2\right)^2\left(1+2\mu^2\right)
                                         +\left(1+\lambda^2\right)^2\mu^4}{\left(1+\lambda^2\right)^2\left(1+\mu^2\right)^2} \,,\\
\\[6pt]
A_{{\bf k}y} &=& J_\perp\frac{1}{1+\lambda^2}+2B\mu\frac{1+2\lambda^2}{\left(1+\lambda^2\right)\left(1+\mu^2\right)}
\\[6pt]
                 &+& 8J_\parallel\frac{\lambda^2\left(1-2\lambda^2\mu^2\right)}{\left(1+\lambda^2\right)^2\left(1+\mu^2\right)^2}\\
\\[6pt]
                 &+&2J_\parallel\gamma_{\bf k}\frac{\left(1+\mu^2\right)^2-\lambda^2\left(1-\mu^2\right)^2}{\left(1+\lambda^2\right)\left(1+\mu^2\right)^2} \,,\\
\\[6pt]
A_{{\bf k}z} &=& J_\perp\frac{1}{1+\lambda^2}+2B\mu\frac{\lambda^2}{\left(1+\lambda^2\right)\left(1+\mu^2\right)}
\\[6pt]
                  &+&8J_\parallel\frac{\lambda^2\left(1+\mu^2\left(1-\lambda^2\right)\right)}{\left(1+\lambda^2\right)^2\left(1+\mu^2\right)^2}
                  +2J_\parallel\gamma_{\bf k}\left(\frac{1-\lambda^2}{1+\lambda^2}\right) \,,\\
\\[6pt]
B_{{\bf k}x} &=& 2J_\parallel\gamma_{\bf k}\frac{\left(1-\lambda^2\right)^2-\left(1+\lambda^2\right)^2\mu^4-8\lambda^2\mu^2}
                                                         {\left(1+\lambda^2\right)^2\left(1+\mu^2\right)^2} \,, \\
\\[6pt]
B_{{\bf k}y} &=& 2J_\parallel\gamma_{\bf k}\frac{\left(1-\mu^4\right)+\lambda^2\left(1-\mu^2\right)^2}
                                                         {\left(1+\lambda^2\right)\left(1+\mu^2\right)^2} \,, \\
\\[6pt]
B_{{\bf k}z} &=& 2J_\parallel\gamma_{\bf k}\frac{\left(1+\lambda^2\right)+\left(1-\lambda^2\right)\mu^4+2\mu^2}
                                                         {\left(1+\lambda^2\right)\left(1+\mu^2\right)^2} \,, \\
\\[6pt]
\frac{C_{\bf k}}{i} &=& B \frac{1-\mu^2}{1+\mu^2} \frac{1}{\sqrt{1+\lambda^2}}
\\[6pt]
                &+& 8J_\parallel\frac{\lambda^2\mu^3}{\left(1+\lambda^2\right)^{3/2}\left(1+\mu^2\right)^2}\left(1-\gamma_{\bf k}\right) \,, \\
\\[6pt]
\frac{D_{\bf k}}{i} &=& -4J_\parallel\gamma_{\bf k}\frac{\mu}{\left(1+\lambda^2\right)^{3/2}\left(1+\mu^2\right)}
                                                   \left(1+\lambda^2\frac{1-\mu^2}{1+\mu^2}\right)
                                                   \,.
\end{eqnarray*}
In the zero field case ($B=0$, $\mu=0$) we have 
$C_{\bf k} = D_{\bf k}= 0$, and therefore $\Omega_{{\bf k}x} = \omega_{{\bf k}x}$,
$\Omega_{{\bf k}y} = \omega_{{\bf k}y}$.
In the FPF phase ($B>B_{c2}$, $\lambda=\infty$, $\mu=1$) we find
$B_{\bf k} = C_{\bf k} = D_{\bf k}= 0$, and the Hamiltonian (\ref{field-fourier-hamiltonian})
is already diagonal in the $\tilde{t}_{{\bf k}\alpha}$ operators.


\end{document}